\documentclass[11pt]{article}
\usepackage{coling2016}
\usepackage{times}
\usepackage{url}
\usepackage{latexsym}
\usepackage{amsfonts}
\usepackage{amsmath}
\usepackage{tabularx}
\usepackage{multirow}
\usepackage{paralist}
\usepackage{graphicx}

\usepackage{algorithm} %format of the algorithm
\usepackage{algorithmic} %format of the algorithm

\title{AttSum: Joint Learning of Focusing and Summarization with Neural Attention}

\author	{
	Ziqiang Cao$^1$ ~~ ~~ Wenjie Li$^1$ ~~ ~~ Sujian Li$^2$ ~~ ~~ Furu Wei$^3$  ~~ ~~ Yanran Li$^1$\\
$^1$Department of Computing, The Hong Kong Polytechnic University, Hong Kong\\
$^2$Key Laboratory of Computational Linguistics, Peking University, MOE, China \\
$^3$Microsoft Research, Beijing, China\\
{\tt \{cszqcao, cswjli, csyli\}@comp.polyu.edu.hk  } \\
{\tt lisujian@pku.edu.cn} \\
{\tt furu@microsoft.com} \\
}

\date{}

\begin{document}
\maketitle
\begin{abstract}
 Query relevance ranking and sentence saliency ranking are the two main tasks in extractive query-focused summarization.  
 Previous supervised summarization systems often perform the two tasks in isolation.
 However, since reference summaries are the trade-off between relevance and saliency, using them as supervision, neither of the two rankers could be trained well.
 This paper proposes a novel summarization system called AttSum, which tackles the two tasks jointly.
 It automatically learns distributed representations for sentences as well as the document cluster.  
 Meanwhile, it applies the attention mechanism to simulate the attentive reading of human behavior when a query is given.
 Extensive experiments are conducted on DUC query-focused summarization benchmark datasets.  
 Without using any hand-crafted features, AttSum achieves competitive performance.
We also observe that the sentences recognized to focus on the query indeed meet the query need. 
\end{abstract}

\section{Introduction}
Query-focused summarization~\cite{dang2005overview} aims to create a brief, well-organized and fluent summary that answers the need of the query.
It is useful in many scenarios like news services and search engines, etc.
Nowadays, most summarization systems are under the extractive framework which directly selects existing sentences to form the summary.
Basically, there are two major tasks in extractive query-focused summarization, i.e., to measure the saliency of a sentence and its relevance to a user's query. 
%A summarization system should select the sentences which both reflect the main ideas of the document cluster and meet the query need to form the summary.
%The first one is to measure the query relevance.
%A query may be a questIion or a task description such as ``Identify and describe types of organized crime that crosses borders or involves more than one country''.
%The generated summary should meet the query need.
%The second task is to measure the sentence saliency.
%In essence, it is still a kind of extractive summarization.
%Therefore it requires to select important sentences to form the summary.
%In extreme cases, some work like \cite{gillick2009scalable} ignore the query information and apply generic summarization approaches to achieve competitive performance.

After a long period of research, learning-based models like Logistic Regression~\cite{li2013using} etc. have become growingly popular in this area.
However, most current supervised summarization systems often perform the two tasks in isolation.
Usually, they design query-dependent features (e.g., query word overlap) to learn the relevance ranking, and query-independent features (e.g., term frequency) to learn the saliency ranking.
Then, the two types of features are combined to train an overall ranking model.
%Then, the two rankers are combined in various ways.
%Errors in the either ranker can seriously worsen the summarization performance.
Note that the only supervision available is the reference summaries.
Humans write summaries with the trade-off between relevance and saliency.
Some salient content may not appear in reference summaries if it fails to respond to the query.
Likewise, the content relevant to the query but not representative of documents will be excluded either.  
%Therefore, reference summaries just act as an intersection of relevant and salient content.
As a result, in an isolated model, weights for neither query-dependent nor query-independent features could be learned well from reference summaries.
%An important sentence may be discarded by the annotator due to its irrelevance, and vice versa.
%With them as supervision, neither separate ranker can be trained well.
%With this training data, the isolated system is unable to learn proper weights for either query-dependent or -independent features.
%For example, in a document cluster with the query ``list the side effects of drugs for mental illness'', there are lots of sentences discuss how to use these drugs properly.
%These sentences are significant enough but irrelevant to the query. 
%The absence of these sentences in the reference summaries largely confuses the weight learning for saliency features.

In addition, when measuring the query relevance, most summarization systems merely make use of surface features like the TF-IDF cosine similarity between a sentence and the query~\cite{wan2009graph}.
However, relevance is not similarity. 
%Even if a sentence is the same as the query (TF-IDF cosine similarity=1), it is useless in the summary because it is unable to answer the query need.
%In paraphrasing, two sentences tell the same thing, which could be measured by text similarity.
%A relevant sentence, however, must be able to meet the query need.
%A summarizer should pick up sentences that are capable of meeting the query need.
%Therefore these surface features are inadequate to measure query relevance. 
Take the document cluster ``d360f'' in DUC\footnote{\url{http://www-nlpir.nist.gov/projects/duc/}} 2005 as an example.
It has the following query: \textit{What are the benefits of drug legalization?}
Here, ``Drug legalization'' are the key words with high TF-IDF scores. 
And yet the main intent of the query is to look for ``benefit'', which is a very general word and does not present in the source text at all.
It is not surprising that when measured by the TF-IDF cosine similarity, the sentences with top scores all contain the words ``drug'' or ``legalization''.
Nevertheless, none of them provides advantages of drug legalization.
See Section~\ref{SC:relevanceExperiment} for reference.
Apparently, even if a sentence is exactly the same as the query, it is still totally useless in the summary because it is unable to answer the query need.
Therefore, the surface features are inadequate to measure the query relevance, which further augments the error of the whole summarization system.
%However, the real intention of the query is to list the advantages, such as the decrease of drug-related accidents.
%Thus, surface features tend to mislead the relevance ranker, which further augments the error of the whole summarization system.
This drawback partially explains why it might achieve acceptable performance to adopt generic summarization models in the query-focused summarization task (e.g., \cite{gillick2009scalable}). 

Intuitively, the isolation problem can be solved with a joint model.
Meanwhile, neural networks have shown to generate better representations than surface features in the summarization task~\cite{cao2015learning,yin2015optimizing}.
%the application of deep learning technologies may work better than surface features to measure query relevance.
Thus, a joint neural network model should be a nice solution to extractive query-focused summarization.
%But how can we design a proper deep neural network?  
%%To handle these two problems, this paper proposes a novel summarization system called AttSum which joints these two tasks in an attention model.
%%AttSum simulates human extractive summarization behavior.
%From a different perspective, let us think about the human behavior for the extractive query-focused summarization.
%When a person attempts to generate the summary according to the query, he will adopt the three steps below:
%\begin{enumerate}
%	\item Measure the semantic matching between the query and a sentence;
%	\item Pay attention on sentences highly related to the query and understand the document meaning;
%	\item Find sentences reflecting the document meaning.
%\end{enumerate}
To this end, we propose a novel summarization system called AttSum, which joints query relevance ranking and sentence saliency ranking with a neural attention model.
The attention mechanism has been successfully applied to learn alignment between various modalities \cite{chorowski2014end,xu2015show,bahdanau2014neural}.
%We use the attention mechanism to automatically measure the query relevance of a sentence.
In addition, the work of \cite{kobayashi-noguchi-yatsuka:2015:EMNLP} demonstrates that it is reasonably good to use the similarity between the sentence embedding and document embedding for saliency measurement, where the document embedding is derived from the sum pooling of sentence embeddings.
%They just use the sum of pre-trained word embeddings to represent sentences and documents.
In order to consider the relevance and saliency simultaneously, we introduce the weighted-sum pooling over sentence embeddings to represent the document, where the weight is the automatically learned query relevance of a sentence.
In this way, the document representation will be biased to the sentence embeddings which match the meaning of both query and documents.
The working mechanism of AttSum is consistent with the way how humans read when having a particular query in their minds.
Naturally, they pay more attention to the sentences that meet the query need.
It is noted that, unlike most previous summarization systems, our model is totally data-driven, i.e., all the features are learned automatically.
%Then, with the query relevance as weights, a sum pooling is followed to combine sentence embeddings to represent the document.
%Finally, we use the similarity of sentence and document embeddings to rank sentences.

We verify AttSum on the widely-used DUC 2005 $\sim$ 2007 query-focused  summarization benchmark datasets.
%In our preliminary experiments, we ignore the Relevance Model and supplement the query information with surface features.
%This simplified approach has attained comparable performance against two popular summarization systems.
%Therefore, we expect the total framework of AttSum can work even better. 
AttSum outperforms widely-used summarization systems which rely on rich hand-crafted features.
We also conduct qualitative analysis for those sentences with large relevance scores to the query.
The result reveals that AttSum indeed focuses on highly query relevant content.

The contributions of our work are as follows:
\begin{itemize}
	\item We apply the attention mechanism that tries to simulate human attentive reading behavior for query-focused summarization;
	\item We propose a joint neural network model to learn query relevance ranking and sentence saliency ranking simultaneously.
	%	\item It achieves competitive summarization performance without any hand-crafted features.
\end{itemize}

\section{Query-Focused Sentence Ranking}
\begin{figure}
	\centering
	\includegraphics[width=0.7\linewidth]{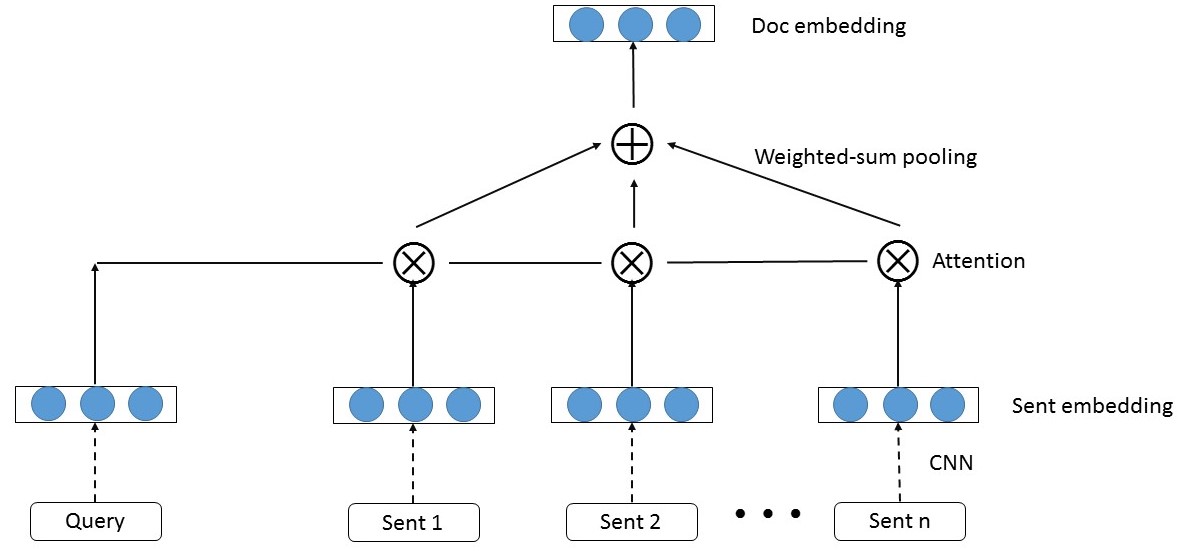}
	\caption{Generation of sentence and document cluster embeddings. ``$\oplus$'' stands for a pooling operation, while ``$\otimes$'' represents a relevance measurement function.}
	\label{fig:AttentionModel}
\end{figure}

%\begin{figure}
%	\centering
%	\includegraphics[width=0.4\linewidth]{../pic/PairwiseRanking}
%	\caption{Pairwise ranking. ``$\odot$'' means a similarity measurement function.}
%	\label{fig:PairwiseRanking}
%\end{figure}

For generic summarization, people read the text with almost equal attention.
However, given a query, people will naturally pay more attention to the query relevant sentences and summarize the main ideas from them.
Similar to human attentive reading behavior, AttSum, the system to be illustrated in this section, ranks the sentences with its focus on the query.
%this section illustrates how AttSum ranks sentences focusing on the query.
The overall framework is shown in Fig.~\ref{fig:AttentionModel}.
From the bottom to up, AttSum is composed of three major layers.
\begin{description}
	\item[CNN Layer] Use Convolutional Neural Networks to project the sentences and queries onto the embeddings.
	\item[Pooling Layer] With the attention mechanism, combine the sentence embeddings to form the document embedding in the same latent space.
	\item[Ranking Layer] Rank a sentence according to the similarity between its embedding and the embedding of the document cluster. 
\end{description}
The rest of this section describes the details of the three layers.

\subsection{CNN Layer}\label{SC:SentenceEmbedding}
Convolutional Neural Networks (CNNs) have been widely used in various Natural Language Processing (NLP) areas including summarization~\cite{cao2015learning,yin2015optimizing}.
They are able to learn the compressed representations of n-grams effectively and tackle the sentences with variable lengths naturally.
We use CNNs to project both sentences and the query onto distributed representations, i.e., 
\begin{align*}
{\bf{v}} (s) &=\text{CNN}(s) \\
{\bf{v}} (q) &=\text{CNN}(q)
\end{align*}

%CNNs have shown promising progress in many NLP areas including summarization ~\cite{cao2015learning,yin2015optimizing}.
%Therefore, AttSum applies CNNs to learn semantic representations for sentences and queries.

%\newcite{Collobert2011} first introduced CNN for a lot of word-related tasks including part-of-speech tagging, chunking, named entity recognition and semantic role labeling.
% Recently, CNN has been applied to model sentences~\cite{kalchbrenner2014convolutional} and achieved excellent results in semantic
%parsing~\cite{yih2014semantic}, search query retrieval~\cite{shen2014learning} as well as relation classification~\cite{zeng2014relation}.
A basic CNN contains a convolution operation on the top of word embeddings, which is followed by a pooling operation.
Let ${\bf{v}}(w_i) \in \mathbb{R}^k$ refer to the $k$-dimensional word embedding corresponding to the $i_{th}$ word in the sentence. 
Assume ${\bf{v}}(w_i : w_{i+j})$ to be the concatenation of word embeddings
$[{\bf{v}}(w_i),\cdots,{\bf{v}}(w_{i+j})]$. 
A convolution operation involves a filter ${\bf{W}}_t^h \in \mathbb{R}^{l \times hk}$, which is applied to a window of $h$ words to produce the abstract features ${\bf{c}}_i^h \in \mathbb{R}^l$:
\begin{equation}\label{EQ:filter}
{\bf{c}}_i^h = f({\bf{W}}_t^h \times {\bf{v}}(w_i : w_{i+j})),
\end{equation}
where $f(\cdot)$ is a non-linear function and  the use of $tanh$ is the common practice. 
To simplify, the bias term is left out.
This filter is applied to each possible window of words in the sentence to produce a feature map.
%\begin{equation}
%C^h = [c_1^h, \cdots , c_{N-h+1}^h]
%\end{equation}
%where $N$ is the length of the sentence. 
Subsequently, a pooling operation is applied over the feature map to obtain the final features $\hat{\bf{c}}^h \in \mathbb{R}^l$ of the filter.
Here we use the max-over-time pooling~\cite{Collobert2011}.
\begin{equation}
\hat{\bf{c}}^h=\max\{{\bf{c}}_1^h, {\bf{c}}_2^h,\cdots\}
\end{equation}
The idea behind it is to capture the most important features in a feature map.
$\hat{\bf{c}}^h$ is the output of CNN Layer, i.e., the embeddings of sentences and queries.

%We apply CNNs to generate embeddings for a sentence $s$ or a query $q$.

%\begin{figure}
%	\centering
%	\includegraphics[width=0.77\linewidth]{pic/CNN}
%	\caption{CNN for sentence/query embedding generation.}
%	\label{fig:CNN}
%\end{figure}

\subsection{Pooling Layer} \label{SC:DocumentEmbedding}
%Then the matching ratio of a sentence to the query is learned in a tensor function.
%Based on the matching ratios, we introduce a weighted-sum pooling operation on sentence embeddings to generate the document embedding which reflects the total document semantics.
%This practice simulates human attention on reading, and makes the document representation automatically biased to query-related sentences.

With the attention mechanism, AttSum uses the weighted-sum pooling over the sentence embeddings to represent the document cluster.
%AttSum firstly learns the vector of a sentence $v(s_i)$, and combines these vectors to represent the cluster $v(D)$.
To achieve this aim, AttSum firstly learns the query relevance of a sentence automatically:
\begin{equation}\label{EQ:tensor}
r(s,q)=\sigma ({\bf{v}}(s) {\bf{M}} {\bf{v}}(q)^T),
\end{equation}
where ${\bf{v}}(s) {\bf{M}} {\bf{v}}(q)^T, {\bf{M}} \in \mathbb{R}^{l \times l}$ is a tensor function, and $\sigma$ stands for the sigmoid function.
The tensor function has the power to measure the interaction between any two elements of sentence and query embeddings.
Therefore, two identical embeddings will have a low score.
This characteristic is exactly what we need. 
To reiterate, relevance is not equivalent to similarity.
%It meets the need that relevance is not similarity.
Then with $r(s,q)$ as weights, we introduce the weighted-sum pooling to calculate the document embedding ${\bf{v}}(d|q)$:
\begin{equation}
{\bf{v}}(d|q) = \sum\nolimits_{s \in d} {r(s,q){\bf{v}}(s)} 
\end{equation}
Notably, a sentence embedding plays two roles, both the pooling item and the pooling weight.
On the one hand, if a sentence is highly related to the query, its pooling weight is large.
On the other hand, if a sentence is salient in the document cluster, its embedding should be representative.
As a result, the weighted-sum pooling generates the document representation which is automatically biased to embeddings of sentences match both documents and the query.  

AttSum simulates human attentive reading behavior, and the attention mechanism in it has actual meaning.
The experiments to be presented in Section~\ref{SC:relevanceExperiment} will demonstrate its strong ability to catch query relevant sentences.
Actually, the attention mechanism has been applied in one-sentence summary generation before~\cite{rush-chopra-weston:2015:EMNLP,hu-chen-zhu:2015:EMNLP}. 
The success of these works, however, heavily depends on the hand-crafted features. 
We believe that the attention mechanism may not be able to play its anticipated role if it is not used appropriately.

\subsection{Ranking Layer} \label{SC:Training}
Since the semantics directly lies in sentence and document embeddings, we rank a sentence according to its embedding similarity to the document cluster, following the work of \cite{kobayashi-noguchi-yatsuka:2015:EMNLP}.
Here we adopt cosine similarity:
\begin{equation}~\label{EQ:similarity}
\cos (d,s|q) = \frac{{{\bf{v}}{{(s)}} \bullet {\bf{v}}(d|q)^T}}{{||{\bf{v}}(s)|| \bullet ||{\bf{v}}(d|q)||}}
\end{equation}
Compared with Euclidean distance, one advantage of cosine similarity is that it is automatically scaled. 
According to~\cite{kaageback2014extractive}, cosine similarity is the best metrics to measure the embedding similarity for summarization.

In the training process, we apply the pairwise ranking strategy~\cite{Collobert2011} to tune model parameters.
Specifically, we calculate the ROUGE-2 scores~\cite{lin2004rouge} of all the sentences in the training dataset.
Those sentences with high ROUGE-2 scores are regarded as positive samples, and the rest as negative samples.
Afterwards, we randomly choose a pair of positive and negative sentences which are denoted as $s^+$ and $s^-$, respectively.
Through the CNN Layer and Pooling Layer, we generate the embeddings of  ${\bf{v}}(s^+)$, ${\bf{v}}(s^-)$ and ${\bf{v}}(d|q)$.
We can then obtain the ranking scores of $s^+$ and $s^-$ according to Eq.~\ref{EQ:similarity}.
With the pairwise ranking criterion, AttSum should give a positive sample a higher score in comparison with a negative sample.
The cost function is defined as follows:
\begin{align}
&\epsilon (d,{s^ + },{s^ - }|q) \\
= &\max (0,\Omega  - \cos (d,{s^ + }|q) + \cos (d,{s^ - }|q)), \notag
\end{align}
where $\Omega$ is a margin threshold.
With this cost function, we can use the gradient descent algorithm to update model parameters.
%The total pairwise ranking process in demonstrated in Fig.~\ref{fig:PairwiseRanking}.
In this paper, we apply the diagonal variant of AdaGrad with mini-batches~\cite{duchi2011adaptive}.
AdaGrad adapts the learning rate for different parameters at different steps.
Thus it is less sensitive to initial parameters than the stochastic gradient descent.

%Two weight matrices, CNN filter $W_n$ and attention weight $M$. 
%\begin{figure}
%	\centering
%	\includegraphics[width=0.66\linewidth]{pic/PairwiseRanking}
%	\caption{Pairwise ranking.}
%	\label{fig:PairwiseRanking}
%\end{figure}

\section{Sentence Selection}\label{sc:selection}
A summary is obliged to offer both informative and non-redundant content.
While AttSum focuses on sentence ranking, it employs a simple greedy algorithm, similar to the MMR strategy~\cite{carbonell1998use}, to select summary sentences.
%The whole process is shown in Algorithm~\ref{alg:greedy}.
At first, we discard sentences less than 8 words like the work of \cite{erkan2004lexrank}.
Then we sort the rest in descending order according to the derived ranking scores.
Finally, we iteratively dequeue the top-ranked sentence, and append it to the current summary if it is non-redundant.
A sentence is considered non-redundant if it contains significantly new bi-grams compared with the current summary content. 
We empirically set the cut-off of the new bi-gram ratio to 0.5.
%A summary is obliged to offer both informative and non-redundant content.
%In this paper, We employ a widely-used greedy method Maximum Marginal Relevance (MMR) \cite{carbonell1998use} to select summary sentences.
%As shown in Algorithm~\ref{alg:mmr}, in each step of selection, the sentence with maximal salience is added into the summary, unless its similarity with a sentence already in the summary exceeds a threshold. Here we use tf-idf cosine similarity and set the threshold $T_{sim}=0.8$.
%\begin{algorithm}[htb]
%	\renewcommand{\algorithmicrequire}{\textbf{Input:}}
%	\renewcommand\algorithmicensure {\textbf{Output:} }
%	\caption{Greedy Sentence Selection Process}
%	\label{alg:greedy}
%	\begin{algorithmic}[1]
%		\REQUIRE ~~\\
%		Sorted sentence array: $s_1,s_2,\cdots,s_N$;\\
%		\ENSURE ~~\\ 
%		Summary: $S$
%		\STATE Initialization: $S=\phi$
%		\FOR{$i=1$; $i\le N$; $i++$}		
%		\IF{length of $s_i \leq 8$ OR bi-gram overlap between $s_i$ and $S \ge 0.5$ } 
%		%		\STATE continue; 
%		%		\ENDIF
%		%		\IF{bi-gram overlap between $s_i$ and $S \ge 0.5$ } 
%		\STATE continue; 
%		\ENDIF
%		\STATE $S = S \cup \{ {s_i}\} $;
%		\IF{length of $S$ reaches the bound}
%		\STATE break;
%		\ENDIF
%		\ENDFOR 
%	\end{algorithmic}
%\end{algorithm}

\section{Experiments}
\subsection{Dataset} \label{SC:dataset}
In this work, we focus on the query-focused multi-document summarization task.
The experiments are conducted on the DUC 2005 $\sim$ 2007 datasets.
All the documents are from news websites and grouped into various thematic clusters.
In each cluster, there are four reference summaries created by NIST assessors. 
We use Stanford CoreNLP\footnote{\url{http://stanfordnlp.github.io/CoreNLP/}} to process the datasets, including sentence splitting and tokenization.
Our summarization model compiles the documents in a cluster into a single document.
Table \ref{TB:dataset} shows the basic information of the three datasets.
We can find that the data sizes of DUC are quite different.
The sentence number of DUC 2007 is only about a half of DUC 2005's. 
For each cluster, a summarization system is requested to generate a summary with the length limit of 250 words. 
We conduct a 3-fold cross-validation on DUC datasets, with two years of data as the training set and one year of data as the test set.
%Since DUC 2003 is query-focused, we use it as the extra development set.

%The model is trained on three years' data and tested on the other year.

\begin{table}[ht]
	\centering
	\begin{tabular}{l|lll}
		\hline
		Year & Clusters & Sentences & Data Source\\ \hline
		2005 & 50       & 45931 &  TREC  \\
		2006 & 59       & 34560  & AQUAINT  \\
		2007 & 30       & 24282   & AQUAINT \\ \hline
	\end{tabular}
	\caption{Statistics of the DUC datasets.}
	\label{TB:dataset}
\end{table} 

\subsection{Model Setting}
For the CNN layer, we introduce a word embedding set which is trained on a large English news corpus ($10^{10}$ tokens) with the  word2vec  model~\cite{mikolov2013efficient}.
The dimension of word embeddings is set to 50, like many previous work (e.g.,  \cite{Collobert2011}).
%In this paper, we adopt the Python implement of word2vec, i.e., gensim\footnote{\url{http://rare-technologies.com/deep-learning-with-word2vec-and-gensim/}}. 
Since the summarization dataset is quite limited, we do not update these word embeddings in the training process, which greatly reduces the model parameters to be learned.
There are two hyper-parameters in our model, i.e., the word window size $h$ and the CNN layer dimension $l$.
We set $h=2$, which is consistent with the ROUGE-2 evaluation.
As for $l$, we explore the change of model performance with $l \in [5,100]$.
Finally, we choose $l=50$ for all the rest experiments.
It is the same dimension as the word embeddings.
During the training of pairwise ranking, we set the margin $\Omega=0.5$.
The initial learning rate is 0.1 and batch size is 100. 

\subsection{Evaluation Metric}
For evaluation, we adopt the widely-used automatic evaluation metric ROUGE~\cite{lin2004rouge}
\footnote{ROUGE-1.5.5 with options: -n 2 -m -u -c 95 -l 250 -x -r 1000 -f A -p 0.5 -t 0}.
It measures the summary quality by counting the overlapping units such as the n-grams, word sequences and word pairs between the peer summary and reference summaries. 
%The parameter of length constraint is ``-l 100'' for DUC 2001/2002, and ``-b 665'' for DUC 2004.  
We take ROUGE-2 as the main measures due to its high capability of evaluating automatic summarization systems~\cite{owczarzak2012assessment}.
%Its recall score is computed as follows:
%\begin{align}
%ROUGE - {2_{{\rm{recall}}}} = \frac{{\sum\limits_{b \in \{ References\} } {{N_{match}}(b)} }}{{\sum\limits_{b \in \{ References\} } {N(b)} }} 
%\end{align}
%where $b$ stands for a bi-gram, and ${N_{match}}(b)$ is the maximum number of bi-grams co-occurring in the peer summary and a set of reference summaries. $N(b)$ is the total number of bi-grams in reference summaries.
%We adopt the recall score as usual.
During the training data of pairwise ranking, we also rank the sentences according to ROUGE-2 scores.

\subsection{Baselines}
To evaluate the summarization performance of AttSum, we implement rich extractive summarization methods.
Above all, we introduce two common baselines.
The first one just selects the leading sentences to form a summary.
It is often used as an official baseline of DUC, and we name it ``LEAD''.
The other system is called ``QUERY\_SIM'', which directly ranks sentences according to its TF-IDF cosine similarity to the query.
In addition, we implement two popular extractive query-focused summarization methods, called MultiMR~\cite{wan2009graph} and SVR~\cite{ouyang2011applying}.
MultiMR is a graph-based manifold ranking method which makes uniform use of the sentence-to-sentence relationships and the sentence-to-query relationships.
SVR extracts both query-dependent and query-independent features and applies Support Vector Regression to learn feature weights.
Note that MultiMR is unsupervised while SVR is supervised.
Since our model is totally data-driven, we introduce a recent summarization system DocEmb~\cite{kobayashi-noguchi-yatsuka:2015:EMNLP} that also just use deep neural network features to rank sentences.
It initially works for generic summarization and we supplement the query information to compute the document representation.

To verify the effectiveness of the joint model, we design a baseline called ISOLATION, which performs saliency ranking and relevance ranking in isolation.
Specifically, it directly uses the sum pooling over sentence embeddings to represent the document cluster.
Therefore, the embedding similarity between a sentence and the document cluster could only measure the sentence saliency.
To include the query information, we supplement the common hand-crafted feature TF-IDF cosine similarity to the query. 
This query-dependent feature, together with the embedding similarity, are used in sentence ranking.
ISOLATION removes the attention mechanism, and mixtures hand-crafted and automatically learned features. 
All these methods adopt the same sentence selection process illustrated in Section ~\ref{sc:selection} for a fair comparison.
%Meanwhile, we implement a state-of-the-art summarization system PriorSum~\cite{cao2015learning} to verify the effectiveness of convolutional neural networks.
%which uses convolutional neural networks to learn document-independent features.
%
%To adapt to the query-focused summarization, we supplement hand-crafted query-dependent features such as word overlap.
%PriorSum is introduced to  

\subsection{Summarization Performance}
The ROUGE scores of the different summarization methods are presented in Table~\ref{TB:Performance}.
We consider ROUGE-2 as the main evaluation metrics, and also provide the ROUGE-1 results as the common practice.
As can be seen, AttSum always enjoys a reasonable increase over ISOLATION, indicating that the joint model indeed takes effects.
With respect to other methods, AttSum largely outperforms two baselines (LEAD and QUERY\_SIM) and the unsupervised neural network model DocEmb.
Although AttSum is totally data-driven, its performance is better than the widely-used summarization systems MultiMR and SVR.
It is noted that SVR heavily depends on hand-crafted features.
Nevertheless, AttSum almost outperforms SVR all the time.
The only exception is DUC 2005 where AttSum is slightly inferior to SVR in terms of ROUGE-2.
Over-fitting is a possible reason.
Table~\ref{TB:dataset} demonstrates the data size of DUC 2005 is highly larger than the other two.
As a result, when using the 3-fold cross-validation, the number of training data for DUC 2005 is the smallest among the three years.
The lack of training data impedes the learning of sentence and document embeddings.

It is interesting that ISOLATION achieves competitive performance but DocEmb works terribly. 
The pre-trained word embeddings seem not to be able to measure the sentence saliency directly.
In comparison, our model can learn the sentence saliency well.

\begin{table}[ht]
	\centering
	\small
	\begin{tabular}{l|lll}
		\hline
		Year                  & Model      & ROUGE-1 & ROUGE-2 \\ \hline
		\multirow{7}{*}{2005} & LEAD       & 29.71   & 4.69    \\
		& QUERY\_SIM & 32.95   & 5.91    \\
		& SVR        & 36.91   & \textbf{7.04}    \\
		& MultiMR    & 35.58   & 6.81    \\ \cline{2-4} 
		& DocEmb     & 30.59   & 4.69    \\
		& ISOLATION  & 35.72   & 6.79    \\
		& AttSum     & \textbf{37.01}   & 6.99    \\ \hline
		\multirow{7}{*}{2006} & LEAD       & 32.61   & 5.71    \\
		& QUERY\_SIM & 35.52   & 7.10    \\
		& SVR        & 39.24   & 8.87    \\
		& MultiMR    & 38.57   & 7.75    \\ \cline{2-4} 
		& DocEmb     & 32.77   & 5.61    \\
		& ISOLATION  & 40.58   & 8.96    \\
		& AttSum     & \textbf{40.90}   & \textbf{9.40}    \\ \hline
		\multirow{7}{*}{2007} & LEAD       & 36.14   & 8.12    \\
		& QUERY\_SIM & 36.32   & 7.94    \\
		& SVR        & 43.42   & 11.10   \\
		& MultiMR    & 41.59   & 9.34    \\ \cline{2-4} 
		& DocEmb     & 33.88   & 6.46    \\
		& ISOLATION  & 42.76   & 10.79   \\
		& AttSum     & \textbf{43.92}   & \textbf{11.55}   \\ \hline
	\end{tabular}
	\caption{ROUGE scores (\%) of different models. We draw a line to distinguish models with or without hand-crafted features. }
	\label{TB:Performance}
\end{table}

%\subsection{Analysis}
\subsection{Query Relevance Performance}\label{SC:relevanceExperiment}
We check the feature weights in SVR and find the query-dependent features hold extremely small weights.
Without these features, the performance of SVR only drops 1\%.
Therefore, SVR fails to learn query relevance well.
The comparison of AttSum and ISOLATION has shown that our method can learn better query relevance than hand-crafted features.
In this section, we perform the qualitative analysis to inspect what AttSum actually catches according to the learned query relevance.
We randomly choose some queries in the test datasets and calculate the relevance scores of sentences according to Eq.~\ref{EQ:tensor}.
We then extract the top ranked sentences and check whether they are able to meet the query need.
Examples for both one-sentence queries and multiple-sentence queries are shown in Table~\ref{TB:query}.
We also give the sentences with top TF-IDF cosine similarity to the query for comparison.

With manual inspection, we find that most query-focused sentences in AttSum can answer the query to a large extent.
For instance, when asked to tell the advantages of drug legalization, AttSum catches the sentences about drug trafficking prevention, the control of marijuana use, and the economic effectiveness, etc.
All these aspects are mentioned in reference summaries.
The sentences with the high TF-IDF similarity, however, are usually short and simply repeat the key words in the query.
The advantage of AttSum over TF-IDF similarity is apparent in query relevance ranking.

When there are multiple sentences in a query, AttSum may only focus on a part of them.
Take the second query in Table~\ref{TB:query} as an example.
Although the responses to all the four query sentences are involved more or less, we can see that AttSum tends to describe the steps of wetland preservation more.
Actually, by inspection, the reference summaries do not treat the query sentences equally either.
For this query, they only tell a little about frustrations during wetland preservation. 
Since AttSum projects a query onto a single embedding, it may augment the bias in reference summaries.
It seems to be hard even for humans to read attentively when there are a number of needs in a query.
Because only a small part of DUC datasets contains such a kind of complex queries, we do not purposely design a special model to handle them in our current work. 
%In comparison, over a half of the sentences with high word overlap to the query just repeat key-phrases.
%Therefore, AttSum is able to automatically learn semantic matching between a query and a sentence.

\begin{table}[]
	\small
	\centering
	\begin{tabular}{p{0.05\linewidth}|p{0.9\linewidth}}
		\hline
		\multirow{4}{*}{AttSum} & It acknowledges that illegal drugs cannot be kept out of the country by tougher border control and interdiction measures.                                                                                              \\ \cline{2-2} 
		& Much greater resources, derived from taxation of the drugs that are now illegal and untaxed and from the billions saved by not wasting money on more criminal- justice measures, must be devoted to drug treatment and drug prevention.                                   \\ \cline{2-2} 
		& As is the case with tobacco, legalizing marijuana, cocaine and heroin would not signify an endorsement of their use.                                                                                                                                                      \\ \cline{2-2} 
		& The consumption and production of marijuana in the United States is on the decrease, and that criminalization costs society more in terms of increased law-enforcement-related costs and deprived revenues from taxes on pot than legalization would.                     \\ \hline
		\multirow{4}{*}{TF-IDF} & Drug prices have soared.                                                                                                                                                                                                                                                  \\ \cline{2-2} 
		& Drug addicts are not welcome.                                                                                                                                                                                                                                             \\ \cline{2-2} 
		& How refreshing to have so much discourse on drugs and legalization.                                                                                                                                                                                                       \\ \cline{2-2} 
		& The only solution now is a controlled policy of drug legalization.                                                                                                                                                                                                        \\ \hline
		Query                   & What are the benefits of drug legalization?                                                                                                                                                                                                                               \\ \hline \hline
		\multirow{4}{*}{AttSum} & Boparai also said that wetlands in many developing countries were vital to the sustenance of human beings, not just flora and fauna.                                                                                                                                      \\ \cline{2-2} 
		& EPA says that all water conservation projects, and agriculture and forestry development along China's major rivers must be assessed in accordance with environmental protection standards, and that no projects will be allowed if they pose a threat to the environment. \\ \cline{2-2} 
		& Finland has agreed to help central China's Hunan Province improve biodiversity protection, environmental education, subtropical forestry and wetlands protection, according to provincial officials.                                                                      \\ \cline{2-2} 
		& The EPA had sought as early 1993 to subject all development on wetlands to strict environmental review, but that approach was rejected by the courts, which ruled in favor of arguments made by developers and by the National Mining Association.                        \\ \hline
		\multirow{4}{*}{TF-IDF} & Statistics on wetlands loss vary widely.                                                                                                                                                                                                                                  \\ \cline{2-2} 
		& Mitigation of any impact on wetlands by creating or enhancing other wetlands.                                                                                                                                                                                             \\ \cline{2-2} 
		& The new regulations would cover about one-fourth of all wetlands.                                                                                                                                                                                                         \\ \cline{2-2} 
		& Now more and more people have recognized wetlands' great ecological and economic potential and the conservation and utilization of wetlands has become an urgent task.                                                                                                    \\ \hline
		Query                   & Why are wetlands important? Where are they threatened? What steps are being taken to preserve them? What frustrations and setbacks have there been?                                                                                                                       \\ \hline
	\end{tabular}
	\caption{Sentences recognized to focus on the query.}
	\label{TB:query}
\end{table}

%\subsubsection{Dimension Impact}\label{SC:DimensionImpact}
%In the previous experiments, we fix the dimension $l$ of the convolutional neural networks to 50, which is the size of the word embeddings.
%Here we inspect the summarization performance with the change of $l$.
%We choose $l \in [10,100]$ and the ROUGE-2 scores of AttSum are demonstrated in Fig.~\ref{fig:ROUGE-dimension}.
%As can be seen, in general, the summarization performance reaches a peak around $l=50$.
%This verifies our practice that the dimension is set to $50$ for all the datasets.
%Notably, the ROUGE-2 scores in DUC 2006 and 2007 show clear properties of the convex function, while in DUC 2005, AttSum works quite steady when $l \leq 60$.  
%One possible reason is that the training set for DUC 2005 is relatively small.
%As a result, too many parameters will lead to serious over-fitting.
%
%\begin{figure}
%	\centering
%	\includegraphics[width=0.7\linewidth]{pic/ROUGE-NumHid}
%	\caption{Changes of ROUGE-2 with the CNN dimension.}
%	\label{fig:ROUGE-dimension}
%\end{figure}

\section{Related Work}
\subsection{Extractive Summarization}
Work on extractive summarization spans a large range of approaches.
Starting with unsupervised methods, one of the widely known approaches is Maximum Marginal Relevance (MMR) \cite{carbonell1998use}. It used a greedy approach to select sentences and considered the trade-off between saliency and redundancy.  
Good results could be achieved by reformulating this as an Integer Linear Programming (ILP) problem which was able to find the optimal solution~\cite{mcdonald2007study,gillick2009scalable}.
Graph-based models played a leading role in the extractive summarization area, due to its ability to reflect various sentence relationships.
For example, \cite{wan2009graph} adopted manifold ranking to make use of the within-document sentence relationships, the cross-document sentence relationships and the sentence-to-query relationships.
In contrast to these unsupervised approaches, there are also various learning-based summarization systems.
Different classifiers have been explored, e.g., conditional random field (CRF)~\cite{galley2006skip}, Support Vector Regression (SVR)~\cite{ouyang2011applying}, and Logistic Regression~\cite{li2013using}, etc.

Many query-focused summarizers are heuristic extensions of generic summarization methods by incorporating the information of the given query.
A variety of query-dependent features were defined to measure the relevance, including TF-IDF cosine similarity~\cite{wan2009graph}, WordNet similarity~\cite{ouyang2011applying}, and word co-occurrence~\cite{prasad2007iiit}, etc.   
However, these features usually reward sentences similar to the query, which fail to meet the query need.

%
%
%LexRank~\cite{erkan2004lexrank} was a popular stochastic graph-based summarization approach, which computed sentence importance grounded on the concept of eigenvector centrality in a graph of sentence similarities. 
%Graph-based methods have a lot of extensions. 
%For example, \cite{wan2008multi} paired graph-based methods with clustering.
%\cite{wan2007manifold,wan2009graph} made uniform use of the sentence-to-sentence relationships and the sentence-to-topic relationships in a manifold-ranking process.
%%\cite{wan2006improved} adds an affinity diffusion process which computes sentence relations with undirected links.
%%Graph-based methods can also be paired with clustering \cite{wan2008multi}.
%
%In contrast to these unsupervised approaches, there are also numerous efforts on supervised learning for summarization where a model is trained to predict the importance. 
%Different classifiers have been explored for this task, such as maximum entropy~\cite{osborne2002using}, conditional random field (CRF)~\cite{galley2006skip}, hidden markov model (HMM)~\cite{conroy2004left} and Support Vector Regression (SVR)~\cite{ouyang2011applying}.
%Besides these metrics which directly modeled sentences, many researches \cite{li2013using,hong2014improving} focused on n-gram regression.
%Recently, treating multi-document summarization as a submodular maximization problem has attracted a lot of interest \cite{lin2011class,sipos2012large,dasgupta2013summarization}.
%
\subsection{Deep Learning in Summarization}
In the summarization area, the application of deep learning techniques has attracted more and more interest. 
\cite{genest2011deep} used unsupervised auto-encoders to represent both manual and system summaries for the task of summary evaluation.
%Their model generated representations of both manual and system summaries. 
%Then, the difference of these two representations were defined as features for regression, with regard to the Pyramid scores~\cite{passonneau2005applying}. 
Their method, however, did not surpass ROUGE.
%Similarly, \cite{SR2015} modeled summarization based on data reconstruction. 
%%They treat a summary as a compressed representation of documents. 
%Using a term frequency vector to represent a sentence, they aimed to find the proper subset which could reproduce the input documents.
%However, their system failed to achieve satisfactory performance.
%Later, \cite{yao2015compressive} improved this idea by introducing an additional sentence dissimilarity term in the optimization framework.
%However, the performance of their summary system is greatly inferior to state-of-the-art approaches like~\cite{wan2014ctsum}.
Recently, some works~\cite{cao2015ranking,cao2015learning} have tried to use neural networks to complement sentence ranking features.
%\cite{cao2015ranking} applied recursive neural networks to automatically learn feature combination, while \cite{cao2015learning} learned document-independent features with convolutional neural networks.
Although these models achieved the state-of-the-art performance, they still heavily relied on hand-crafted features.
A few researches explored to directly measure similarity based on distributed representations.
\cite{yin2015optimizing} trained a language model based on convolutional neural networks to project sentences onto distributed representations.
\cite{cheng2016neural} treated single document summarization as a sequence labeling task and modeled it by the recurrent neural networks.
Others like~\cite{kobayashi-noguchi-yatsuka:2015:EMNLP,kaageback2014extractive} just used the sum of trained word embeddings to represent sentences or documents.

In addition to extractive summarization, deep learning technologies have also been applied to compressive and abstractive summarization.
\cite{filippova-EtAl:2015:EMNLP} used word embeddings and Long Short Term Memory models (LSTMs) to output readable and informative sentence compressions.
\cite{rush-chopra-weston:2015:EMNLP,hu-chen-zhu:2015:EMNLP} leveraged the neural attention model~\cite{bahdanau2014neural} in the machine translation area to generate one-sentence summaries.
We have described these methods in Section~\ref{SC:DocumentEmbedding}.

\section{Conclusion and Future Work}
This paper proposes a novel query-focused summarization system called AttSum which jointly handles saliency ranking and relevance ranking.
It automatically generates distributed representations for sentences as well as the document cluster. 
Meanwhile, it applies the attention mechanism that tries to simulate human attentive reading behavior when a query is given.
%AttSum simulates the human summarization behavior that naturally pays more attention on sentences related to the query.
% Specifically, we use convolutional neural networks to generate embeddings for both sentences and the query.
% Then the matching ratio of a sentence to the query is measured according to their embeddings.
% Based on matching ratios, we introduce a weighted-sum pooling operation on sentence embeddings to generate the document embedding which reflects the summary semantics.
% Finally, we select sentences best reproducing the document embedding to form the summary.
We conduct extensive experiments on DUC query-focused summarization datasets.
Using no hand-crafted features, AttSum achieves competitive performance.
It is also observed that the sentences recognized to focus on the query indeed meet the query need.

Since we have obtained the semantic representations for the document cluster, we believe our system can be easily extended into abstractive summarization.
The only additional step is to integrate a neural language model after document embeddings.
We leave this as our future work.

%\section*{Acknowledgements}
%The work described in this paper was supported by Research Grants Council of Hong Kong (PolyU 152094/14E, PolyU 152248/16E), National Natural Science Foundation of China (61272291 and 61672445) and The Hong Kong Polytechnic University (4-BCB5, B-Q46C and  G-YBJP).
%The correspondence authors of this paper are Wenjie Li and Sujian Li.

\bibliographystyle{acl}
\bibliography{ijcai2016}

\end{document}